\documentclass[prb, amsmath,reprint, groupedaddress, showpacs]{revtex4-1}
\usepackage{amsmath,amssymb,pdfpages,bm,amsthm, epstopdf}

\usepackage{amsfonts}
\usepackage{graphicx}
\usepackage{mathtools}
\usepackage{fullpage}
\usepackage{url}

\newcommand{\e}{\mathrm{e}}
\newcommand{\dd}{\; \mathrm{d}}
\newcommand{\fA}{\mathcal{A}}
\newcommand{\fB}{\mathcal{B}}
\newcommand{\veps}{\varepsilon}
\newcommand{\vf}{v_{F}}
\begin{document}

\title{Klein tunneling and cone transport in AA-stacked bilayer graphene}
\author{Matthew Sanderson and Yee Sin Ang}
\affiliation{School of Physics, University of Wollongong,  NSW 2522, Australia}
\author{C. Zhang}
\email{czhang@uow.edu.au}
\affiliation{School of Physics and Institute for Superconducting and Electronic Materials, University of Wollongong, NSW 2522, Australia}

\begin{abstract}
We investigate the quantum tunneling of electrons in an AA-stacked bilayer graphene (BLG) $n$-$p$ junction and $n$-$p$-$n$ junction. 
We show that Klein tunneling of an electron can occur in this system. 
The quasiparticles are not only chiral but are additionally described by a `cone index'. 
Due to the orthogonality of states with different cone indexes, electron transport across a potential barrier must strictly conserve the cone index and this leads to the protected cone transport which is unique in AA-stacked BLG. 
Together with the negative refraction of electrons, electrons residing in different cones can be spatially separated according to their cone index when transmitted across an $n$-$p$ junction. 
This suggests the possibility of `cone-tronic' devices based on AA-stacked BLG.
Finally, we calculate the junction conductance of the system.
\end{abstract}

\maketitle
\section{Introduction}
Since the isolation of single layers of graphite in 2003
\cite{novo1}, many exciting works on single layer graphene (SLG)
have been done \cite{novorev}. For example, the prediction and
observation of electron-hole symmetry and a half-integer quantum
Hall effect \cite{novo2,zhang1,berg},  finite conductivity at zero
charge-carrier concentration \cite{novo2}, the strong suppression of
weak localization \cite{suzuura,morozov}, universal conductance
\cite{gus,kuz,nair,chao-ma}, magnetic enhancement of optical
conductance in graphene nanoribbons \cite{liu1} and strong nonlinear optical
response in the terahertz frequency regime\cite{mikh,arw09}.

Bilayer graphene (BLG) exhibits additional new properties not seen
in single layer graphene, chief among them are the trigonal
warping\cite{mccann_prl, trig_effect}, a phenomenon solely due to
the interlayer coupling. The trigonal warping is robust and
independent of the interlayer coupling strength. The quantum Hall
plateaus in BLG are doubled and are also independent of the interlayer
coupling strength\cite{QHE}. In general, electrons in BLG
behave qualitatively differently than in SLG. Phenomena
such as interlayer drags\cite{taka93} and correlations\cite{zhang94}
are unique in BLG. The electronic and transport properties of
BLG differ significantly from SLG in many respects, particularly at the low energy `Dirac' regime. Various models for low energy
BLG exist in the literature depending on the coupling terms
included, and whether electronic bands beyond the lowest energy
subbands are retained \cite{nilscast, mccann_prl}. Many interesting
results are obtained based on a model that includes only the most
dominant of the interlayer coupling terms in BLG, as well as the
usual nearest neighbor intralayer term \cite{castro1}. By including
the second most dominant interlayer coupling, some unusual
properties such as a peculiar Landau-level spectrum \cite{mccann_prl}, a new low energy peak in the
optical conductance \cite{abe,nicol} and retro-reflection of electrons in a BLG/superconductor heterojunction \cite{Ang} have been
demonstrated. By further increasing the number of layers, one has graphene multilayers whose energy dispersion
near the $K$-point can be tuned by a gate voltage\cite{peeters}.

BLG has two distinct forms: the usual Bernal stacking (or AB-stacking) where the A-sublattice of the top layer is stacked directly above the B-sublattice of the bottom layer, and the AA-stacking (Fig. 1) where one layer is stacked directly above the other with exact matching of the sublattices \cite{bilayer_electronic_properties, AA_conductivity}. In AA-stacked BLG, the energy dispersion in the low energy regime is made up of two shifted Dirac cones. The Dirac cones in AA-stacked BLG are identical to that of the linear energy dispersion in SLG, except that the two cones are offset from the SLG dispersion by the interlayer hopping energy. There has been considerably less research into AA-stacked BLG than AB-stacked BLG as it was believed to be unstable. However recent studies have shown that BLG in the form of AA-stacking can actually be formed stably \cite{AA_stack_exp,graphene_stacking,graphene_covalent_bonds}. 

\begin{figure}
	\includegraphics[width=4cm]{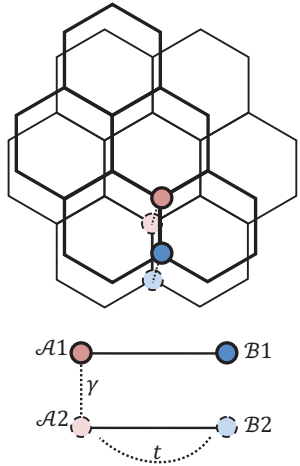}
	\caption{(Color online). Lattice structure of AA-stacked bilayer graphene. The $\fA1$ sublattice is stacked directly above the $\fA2$ sublattice. The interlayer and intralayer hopping energies are $\gamma$ and $t$ respectively.}
	\label{fig:finite_width_junction}
\end{figure}

The Klein paradox is an ultra-relativistic phenomenon where massless Dirac fermions transmit perfectly across a potential barrier irrespective of the height and width of the barrier \cite{Klein}. In the ultra-relativistic case, the transmitted electronic wavefunction does not exponentially decay while propagating within the potential barrier. Instead, the incident electron state is connected to the positron state within the barrier which allows the undamped propagation of the incident particle across the potential barrier. A condensed matter version of the celebrated Klein paradox has been demonstrated in SLG and AB-stacked BLG where the low energy electron is described by relativistic massless and massive Dirac fermion, respectively \cite{klein_tunneling_multiple_barriers,chiral_tunneling,klein_tunneling}. Due to the conservation of pseudospin, perfect electron transmission occurs in SLG at normal incidence. In contrast to SLG, the pseudospin of electrons in AB-stacked BLG rotates twice as fast as the electron wavevector and this results in the perfect reflection of electrons at normal incidence. This peculiar `Klein reflection' has no counterpart in relativistic quantum mechanics. Cloaking of electronic states in AB-stacked BLG based on `Klein reflection' has been proposed \cite{Gu}. 

In this work, we study theoretically the electron transport across a junction in AA-stacked BLG. For AA-stacked BLG, electron transport across a magnetoelectric barrier has been considered previously \cite{AA_stack_transport_mag_barrier}. Here we shall study the problem of electron tunneling across a non-magnetoelectric step potential barrier ($n$-$p$ junction and $n$-$p$-$n$ junction) in AA-stacked BLG. We found that Klein tunneling occurs at normal incidence as a result of the Dirac nature of the quasiparticles. Furthermore, the quasiparticles in AA-stacked BLG are described by a `cone index' and we show that it is a strictly conserved quantity. Such conservation leads to protected cone transport of electrons in AA-stacked BLG which could potentially have applications in a device. Finally, we calculate the junction conductance and the results are discussed.

The paper is organized as followed. In Section II we present the electronic properties of AA-stacked BLG. Electron transport across $n$-$p$ and $n$-$p$-$n$ junctions are presented in section III. The tunneling conductance is presented in section IV, followed by a conclusion in section V. Additional derivations and discussions are given in the Appendix.

\section{Tight binding Hamiltonian of AA-stacked BLG}

\begin{figure*}
	\centering
	\includegraphics[width=15cm]{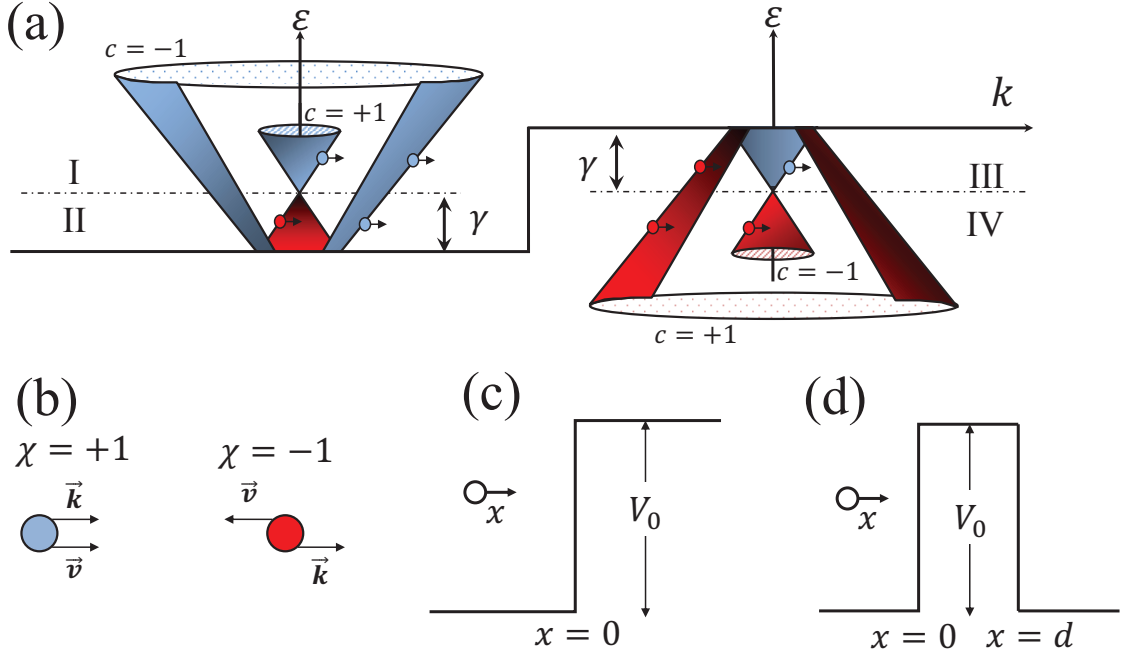}
	\caption{(Color online). (a) Band structures of electrons (left) and holes (right). The band structure is made up of two up-down shifted Dirac cones. Electrons and holes reside in one of the Dirac cones which is labeled by a cone index $c$. (b) The chirality of the quasiparticles are labeled by $\chi$. For $\chi=+1$($\chi = -1$), the wavevector is (anti-)parallel to the group velocity. (c) and (d) show the tunneling of an electron through an $n$-$p$ junction and an $n$-$p$-$n$ junction respectively}.
	\label{fig:AA_structure}
\end{figure*}

The Hamiltonian of AA-stacked BLG is formulated by considering a tight-binding model \cite{tight_binding}. Only nearest neighbor in-plane hopping is considered, as well as direct interlayer hopping (ie. \!\!$\fA$ to $\fA$ or $\fB$ to $\fB$ sublattices). The Hamiltonian takes the form \cite{AA_stack_instability}:
\begin{widetext}
\begin{equation}
H = -t\!\!\!\sum_{\langle \mathbf{nm}\rangle i\sigma}{a^{\dagger}_{\mathbf{n} i\sigma}b_{\mathbf{m} i\sigma}} - \gamma\sum_{\mathbf{n}\sigma}{a^{\dagger}_{\mathbf{n}1\sigma}a_{\mathbf{n}2\sigma}} - \gamma\sum_{\mathbf{m}\sigma}{b^{\dagger}_{\mathbf{m}1\sigma}b_{\mathbf{m}2\sigma}} + H.C.,
\end{equation}
\end{widetext}
where \noindent $a^{\dagger}_{\mathbf{n} i\sigma}$ ($a_{\mathbf{n} i\sigma}$) and $b^{\dagger}_{\mathbf{m} i\sigma}$ ($b_{\mathbf{m} i\sigma}$) are the creation (annihilation) operators on the $\fA$ and $\fB$ sublattices respectively for layer $i = 1,2$ and spin $\sigma$. In the basis of $\psi = (a_{1},b_{2},a_{2},b_{1})^{T}$, the Hamiltonian operator is written as\cite{AA_conductivity}
\begin{equation}
\hat{H} =
\begin{pmatrix}
0				&	0				&	\gamma		&	f(\mathbf{k})	\\
0				&	0				&	\overline{f}(\mathbf{k})	&	\gamma	\\
\gamma	&	f(\mathbf{k})	&	0		&	0					\\
\overline{f}(\mathbf{k}) & \gamma &	0	&	0	\\
\end{pmatrix}.
\label{eq:matrix_Hamiltonian}
\end{equation}
Here, $\gamma\approx 0.2$ eV is the interlayer hopping energy\cite{AA_conductivity}, $f(\mathbf{k}) = \hbar \vf k \e^{-i\theta}$ where $\theta$ is the angle of the wavevector, $\vf = \frac{\sqrt{3}ta}{2\hbar} \approx 10^{6}$ m/s is the Fermi velocity, $t$ is the in-plane nearest neighbor hopping energy, and $a$ is the in-plane carbon-carbon separation. The Hamiltonian can be compactly written as:
\begin{equation}
\hat{H} = \sigma_{x}\otimes(\gamma \hat{I}_{2} + \boldsymbol\sigma\cdot\mathbf{k}).
\end{equation}
where $\hat{I}_{2}$ is the $2\times 2$ identity matrix and $\mathbf{\sigma}$ is the Pauli spin matrix. The Hamiltonian in Eq.~\eqref{eq:matrix_Hamiltonian} can be diagonalized to give the energy dispersion $E_{\eta_{1}\eta_{2}}(\mathbf{k}) = \eta_{1}\left(|f(\mathbf{k})| + \eta_{2}\gamma\right) = \eta_{1}\left(\hbar v_{f} |\mathbf{k}| + \eta_{2}\gamma\right)$, where $\eta_{1} = \pm 1$ and $\eta_{2} = \pm 1$ are the band indices. The band structure of AA-stacked BLG is shown in Fig. 2a. The band structure is made up of two up-down shifted Dirac cones. For an intrinsic AA-stacked BLG, the Dirac cones intersect each other at $E=0$. Instead of forming a single Dirac point there, a Dirac `ring' is formed. The eigenvectors are given by
\begin{equation}
\label{eq:wavefunction}
\psi_{\eta_{1}\eta_{2}}(\theta) = \e^{i\mathbf{k}\cdot\mathbf{r}}
\begin{pmatrix}
\eta_{1} \e^{-i\theta} \\
\eta_{1}\eta_{2} \\
\eta_{2} \e^{-i\theta} \\
1
\end{pmatrix}.
\end{equation}
We will suppress the normalization factor throughout this paper as it will have no impact on the final results.
The wavevector can be decomposed into its $x$ and $y$ components given by $\mathbf{k} = \left(k_{x}, k_{y}\right) = |\mathbf{k}|\left(\cos\theta,\sin\theta\right) = (\hbar v_{F})^{-1}\left[\eta_{1} E_{\eta_{1}\eta_{2}} - \eta_{2}\gamma\right]\left(\cos\theta,\sin\theta\right)$.

We define two operators that commute with the Hamiltonian:
\begin{subequations}
	\begin{align}
	\hat{C} = \gamma^{5} &= \sigma_{x}\otimes\hat{I}_{2} = 
	\begin{pmatrix}
		0 & 0 & 1 & 0 \\
		0 & 0 & 0 & 1 \\
		1 & 0 & 0 & 0 \\
		0 & 1 & 0 & 0
	\end{pmatrix},\\
	\hat{\chi} &= \sigma_{x}\otimes\frac{\left(\boldsymbol\sigma\cdot\mathbf{k}\right)}{k},
	\end{align}
\end{subequations}
where $\gamma^{5}$ is the $5^{th}$ gamma matrix. Both of these operators are the Kronecker product of $\sigma_{x}$ with an operator that commutes with the Hamiltonian for SLG (i.e. the identity matrix $\hat{I}_{2}$ and the chirality operator for SLG $\frac{\hat{\boldsymbol\sigma}\cdot\mathbf{k}}{k}$). We shall refer to the operators $\hat{C}$ and $\hat{\chi}$ as the `cone operator' and `chirality operator' respectively. We define the following terms: $c\equiv\eta_1\eta_2$ and $\eta_1 \equiv\chi$ where $c$ will be referred to as the `cone index' and $\chi$ the `chirality index'. The energy dispersion and the eigenfunctions can be recast in terms of $c$ and $\chi$, as follows:
\begin{subequations}
\begin{align}
E_{c,\chi}(\mathbf{k}) = \chi \hbar v_{f} |\mathbf{k}| + c\gamma \\
\psi_{c,\chi}(\theta) = \e^{i\mathbf{k}\cdot\mathbf{r}}
\begin{pmatrix}
\chi \e^{-i\theta} \\
c \\
c\chi \e^{-i\theta} \\
1
\end{pmatrix}.
\end{align}
\end{subequations}
Using these definitions, it can be shown that $\hat{C}\psi_{c,\chi} = c \psi_{c,\chi}$ and $\hat{\chi}\psi_{c,\chi} = \chi \psi_{c,\chi}$. The energy bands (see Fig. 2a) can be conveniently labeled by the cone and chirality indexes. 
A quasiparticle state is situated in the upper (lower) cone if $c=+1$ ($c=-1$). The physical significance of the eigenvalue $\chi$ is immediately obvious if we consider the group velocity \cite{theoretical_graphene};
\begin{equation}
\mathbf{v}_{\chi}(\mathbf{k}) = \chi v_{F}\frac{\mathbf{k}}{k}.
\label{eq:velocity}
\end{equation}
Whether $\mathbf{v}(\mathbf{k})$ aligns with $\mathbf{k}$ is dependent on the chirality index $\chi$. 
A quasiparticle state is electron-like (i.e. $\mathbf{v}(\mathbf{k})$ is parallel with $\mathbf{k}$) if $\chi=+1$, and hole-like (i.e. $\mathbf{v}$ is anti-parallel with $\mathbf{k}$) if $\chi=-1$ (Fig. 2b), hence the naming of $\chi$ as the `chirality index' of a quasiparticle. In Fig. 2a, it can be seen that the electron states can be subdivided into two regimes: (I) where both $c=\pm 1$ electrons are electron-like and (II) where $c=+1$ electrons are hole-like while $c=-1$ electrons are electron-like. The holes can also have two chiralities as denoted by Regimes (III) and (IV). 

\section{Klein tunneling through a potential barrier}

\subsection{n-p junction}
We first consider electron tunneling through an $n$-$p$ junction (see Fig. 2c). In graphene, one can use a gate voltage to control the Fermi energy to form the $p$ and $n$ regions. The band discontinuity at x=0 gives rise to a step-like potential barrier having the form
\begin{equation}
V(x) =
\begin{dcases}
V_{0} & \text{when } x > 0\\
0 & \text{when } x < 0.
\end{dcases}
\end{equation}
As we shall show below, all inter-cone transitions (i.e. $c\to -c$ processes) are strictly forbidden due to the orthogonality of electron wavefunctions with a different cone index. For the permissible intra-cone processes, there exist four cases for transitions across the barrier: (i) electron in Regime I $\to$ hole in Regime III; (ii) electron in Regime I $\to$ hole in Regime IV; (iii) electron in Regime II $\to$ hole in Regime III; and (iv) electron in Regime II $\to$ hole in Regime IV. The transmission coefficient of each of the four processes can be calculated using the same method. First we shall show in detail the calculation for the transmission of an electron in case (ii), i.e. electron with $c=+1$ and $\chi=-1$ into a transmitted state with the same $c$ and $\chi$ as the incident state. In this case, $E < \gamma < V$ is the incident electron energy. The results for the other possibilities all follow analogously. 

The wavefunctions on the left-hand side and right-hand side of the barrier, $\Psi_L$ and $\Psi_R$ respectively, can be written using Eq.~\eqref{eq:wavefunction} as:
\begin{subequations}
	\begin{align}
	\psi_{L} = \e^{ik_{x}x}
	\begin{pmatrix}
		-\e^{-i\phi} \\ 1 \\ -\e^{-i\phi} \\ 1
	\end{pmatrix}
	+ r\e^{-ik_{x}x}
	\begin{pmatrix}
		\e^{i\phi} \\ 1 \\ \e^{i\phi} \\ 1
	\end{pmatrix}\allowdisplaybreaks[4]\\
	\psi_{R} = t\e^{ik'_{x}x}
	\begin{pmatrix}
		-\e^{-i\theta_{t}} \\ 1 \\ -\e^{-i\theta_{t}} \\ 1
	\end{pmatrix}.
	\end{align}
\end{subequations}
Here $\phi$ is the angle of the incident electron relative to the normal to the barrier, $\theta_{t}$ is the transmitted angle and $k_{x}$ and $k'_{x}$ are the x-components of the wavevector for the incident and transmitted electrons respectively. 
A factor of $e^{ik_{y}y}$ has been omitted for simplicity. 
$\theta_{t}$ and $k'_{x}$, can be written explicitly as:
\begin{align}
	\theta_{t} &= \arcsin\left[\frac{1-\varepsilon}{1-\varepsilon+\nu}\sin\phi\right] \label{eq:theta_t_eg}\\
	k'_{x} &= \frac{\gamma}{\hbar v_{f}}\sqrt{\nu^{2} -2\nu(\varepsilon-1) + (\varepsilon-1)^{2}\cos^{2}\phi} \label{eq:kx_eg},
\end{align}
where $\nu = V/\gamma$ and $\varepsilon = E/\gamma$ are the dimensionless potential and incident electron energy respectively. It must be remembered that as the energy and potential terms are being parameterised by $\gamma$, $\gamma \to 1$ in this non-dimensionalised form, so whenever a sum appears of the form $\veps + c$ etc. the result obtained by re-introducing the dimensions is $\frac{1}{\gamma}(E + c\gamma)$.

There exists a critical incident angle given by Eq.~\eqref{eq:crit_angle} at which the transmitted angle is $\pi/2$. When $\phi>\phi_{crit}$, transmission state is evanescent and the electron must be reflected.
\begin{equation}
	\phi_{crit} = \arcsin\left[1+\frac{\nu}{1-\varepsilon}\right].
	\label{eq:crit_angle}
\end{equation}
For $c=+1$ incident states, there is no critical angle in the range of $E < V/2 + \gamma$. When $E > V/2 + \gamma$, the Fermi radius of the incident electron in $k$-space is larger than that of the transmitted hole and the critical angle becomes relevant. When the incident angle $\phi>\phi_{crit}$, the transmitted mode is evanescent and the electron must be reflected as in the case of total internal reflection in optics. For a $c=-1$ incident electron, this occurs at the incident energy $E > V/2 - \gamma$.

The expressions for transmitted angle (Eq.~\eqref{eq:theta_t_eg}), transmitted wavevector (Eq.~\eqref{eq:kx_eg}) and critical angle (Eq.~\eqref{eq:crit_angle}) can be respectively written in generality:

\begin{align}
\theta_{t} &= \arcsin\left[\mu\frac{\veps - c}{\veps - c - \nu}\sin\phi\right] \label{eq:theta_t_gen} \\
k'_{x} &= \chi'\frac{\gamma}{\hbar\vf}\text{sgn}(\veps - c - \nu) \nonumber \\
&\times \sqrt{\nu^{2} - 2\nu(\veps - c) + (\veps - c)^{2}\cos^{2}\phi} \label{eq:kx_gen} \\
\phi_{crit} &= \arcsin\left[\mu\frac{\veps - c - \nu}{\veps - c}\right] \label{eq:crit_gen}
\end{align}
where $\mu = \chi\chi'$ is the product of chirality indexes.

The transmission probability for the intra-cone, i.e. $c\to c$ process, is found to be
\begin{equation}
	\label{eq:trans}
	T = \frac{\cos\phi\cos\theta_{t}}{\cos^{2}\left(\frac{\phi+\theta_{t}}{2}\right)}.
\end{equation}

\begin{figure}
	\includegraphics[width=7.5cm]{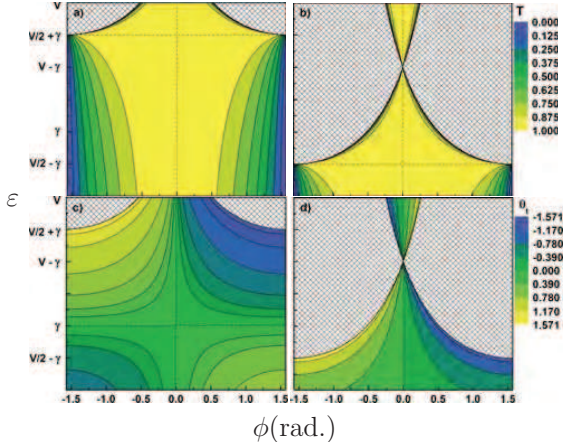}
	\caption{(Color online). Transmission probabilities for the (a) upper and (b) lower cones across an $n$-$p$ junction. The transmission angle is shown in (c) for the upper cone and in (d) for the lower cone. Cross-hatched pattern represents an evanescent transmitted mode}
	\label{fig:ul_plots}
\end{figure}

The transmission probabilities for intra-cone process of $c\to-c$ is zero. This can be understood by considering the step potential as a sudden perturbation. In the presence of a sudden approximation, the probability for an initial wavefunction $\psi_{c,\chi}(\theta)$ to undergo a sudden transition into a new state of $\psi_{c',\chi'}(\theta')$ is given by $P = \left|\langle\psi_{c,\chi}(\theta)^{\dagger}|\psi'_{c',\chi'}(\theta')\rangle\right|^{2}$. It can be shown that
\begin{equation}
\langle\psi_{c,\chi}(\theta)^{\dagger}|\psi'_{c',\chi'}(\theta')\rangle = (1 + cc')(\chi\chi'\e^{i\varphi}+1)
\label{eq:psi_ortho}
\end{equation}
where $\varphi = \theta-\theta'$. This immediately shows that if $c\neq c'$ (i.e. $cc'=-1$), the transition probability is strictly zero ($P=0$) \cite{roldan}. The cone index is therefore a conserved quantity in the presence of a sudden perturbation. The product of the chirality indices $\chi\chi'$ in general does not give rise to $P=0$ when $\chi \neq \chi'$. Therefore, $c$ is a strictly conserved quantity while $\chi$ is not necessarily conserved during a state transition. The orthogonality of the wavefunction leads to the conserved cone transport in AA-stacked BLG across a potential barrier; the electron is constrained to move within the same cone across a junction \cite{AA_conductivity}. This cone transport however does not occur when a magnetoelectric barrier is present \cite{AA_stack_transport_mag_barrier}. The cone transport in AA-stacked BLG highlights a major difference between its Dirac fermions and those in SLG: the AA-BLG quasiparticles are not only described by pseudospin as in the case of SLG, but are also labeled by an additional cone index, $c$.

The transmission probabilities of $c=+1$ and $c=-1$ incident electrons are plotted in Fig.~\ref{fig:ul_plots}a and Fig.~\ref{fig:ul_plots}b respectively. The vertical and horizontal dashed lines indicates the Klein tunneling effect where the probability of transmission is 100\%. As the band structure of AA-stacked BLG is composed of two shifted Dirac cones an additional cone index is introduced. The chirality of the massless Dirac quasiparticle however is not altered, therefore Klein tunneling is allowed to occur in AA-stacked BLG.

\begin{figure}
	\includegraphics[width=4.5cm]{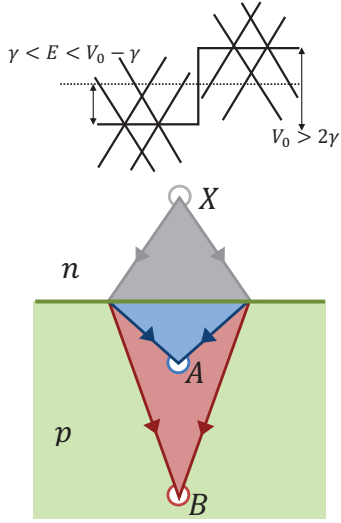}
	\caption{(Color online). Schematic drawing of a cone polarizer based on AA-stacked BLG $n$-$p$ junction: (a) Incident energy potential height to achieve cone polarization. The incident energy has to be in the range of $\gamma < E < V_0 - \gamma$ and the potential height has to be greater than $2\gamma$. (b) Spatial separation of electrons according to their cone index. Electrons injected at point X are refracted through different transmission angles according to their cone index. $c=+1$ electrons are focused at Point B while $c=-1$ electrons are focused at Point A.}
	\label{cone}
\end{figure}

The transmitted angles are plotted in Fig.~\ref{fig:ul_plots}c and Fig.~\ref{fig:ul_plots}d.
The transmission of electrons can be thought of analogously to the usual refraction of a light ray across the interface of two media of different refractive index. If the incident angle and the transmitted angle has the same sign, we have the usual refraction. If the incident angle and the transmitted angle have opposite signs the electron transmission is equivalent to the optical case of negative refraction.
The horizontal dashed line in Fig. 3c indicates the transition of a $c=+1$ electron from usual refraction into negative refraction. The transition occurs because the incident electron becomes hole-like ($\chi=-1$) when the incident energy is lower than $\gamma$. Due to this, the $c=+1$ electrons can either be focused or be divergently refracted depending on its incident energy. In Fig. 3d, the chirality of the incident $c=-1$ electron is always $\chi=+1$. The transmitted hole however can be either of $\chi=\pm1$, therefore the transmission can be the usual refraction or the negative refraction depending on whether the incident energy is greater than or less than $V-\gamma$, as seen in Fig. 3d.

The protected cone transport and the negative refraction of electrons across the AA-stacked BLG $n$-$p$ junction can be used to create a `cone polarizer'. We consider a specific case when the incident electron energy is $V-\gamma>E>\gamma$ and the potential height is $V>2\gamma$ (as indicated in Fig. \ref{cone}a). In this case, both $c=\pm 1$ incident electrons have a chirality of $\chi = +1$ (i.e. they are electron-like and the electron motion is parallel with the wavevector). Upon transmission, incident states residing in the above mentioned energy regime are transmitted into hole states of chirality $\chi=-1$. The transmission involving a flip of the chirality index is in the same case as the negative refraction of massless Dirac fermion in an $n$-$p$ junction in single layer graphene \cite{cheianov}. For electrons injected from a point source, the negatively refracted electrons are focused behind the junction \cite{cheianov} as is analogous to the light focusing effect of a negative refractive index medium \cite{pendry}. From conservation of wavevector component parallel to the junction, the transmitted angles can be written as:
\begin{equation}
\sin{\theta_t^{(c)}} = \frac{k_i^{(c)}}{k_t^{(c)}} \sin{\phi}
\end{equation}
where $k_i^{(c)}$ and $k_t^{(c)}$ are the incident and transmitted wavevectors respectively for electrons in the cone denoted by $c$. This immediately shows that the magnitude of $\theta_t^{(c)}$ is dependent on the Fermi radii of the incident and transmitted states for a given incident angle $\phi$. For $c=+1$ incident electrons, $k_i^{(+)}<k_t^{(+)}$, while for $c=-1$ incident electrons, $k_i^{(-)}<k_t^{(-)}$, therefore $\theta_t^{(+)}<\theta_t^{(-)}$. The $c=+1$ electrons are weakly refracted through a smaller angle in comparison to the case of $c=-1$ electron. As illustrated in Fig. \ref{cone}b, the $c=+1$ electrons injected from a point source are therefore focused at a position further away from the junction (at Point B in Fig. \ref{cone}b). In contrast, the $c=-1$ electrons are strongly refracted through a larger angle, and hence are focused at a position closer to the junction (at Point B in Fig. \ref{cone}b). Consequently, the $n$-$p$ junction, operating in the energy regime of $V_0-\gamma>E>\gamma$ and the potential height is $V_0>2\gamma$, is essentially a `cone polarizer' which spatially resolves electrons of different cone index upon transmission.

\subsection{n-p-n junction}
We now consider the tunneling of electron through a finite-width potential barrier (Fig. 2d):
\begin{equation}
V(x) =
\begin{dcases}
V_{0} & \text{when } 0 < x < d\\
0 & \text{otherwise}
\end{dcases}
\end{equation}
where the potential width is $d$. The total wavefunction in each region can be written as,
\begin{equation}
\psi_{L} = \psi_{i} + \psi_{r}, \quad \psi_{C} = \psi_{a} + \psi_{b}, \quad \psi_{R} = \psi_{t},
\end{equation}
where,
\begin{subequations}
	\begin{align}
		\psi_{i} &= \e^{ik_{x}x}
		\begin{pmatrix}
			\chi \e^{-i\phi} \\ c\\ \chi c \e^{-i\phi} \\ 1
		\end{pmatrix}\allowdisplaybreaks[4] \\
		\psi_{r} &= r\e^{-ik_{x}x}
		\begin{pmatrix}
			-\chi \e^{i\phi} \\ c \\ -\chi c \e^{i\phi} \\ 1
		\end{pmatrix}\allowdisplaybreaks[4] \\
		\psi_{a} &= a\e^{ik'_{x}x}
		\begin{pmatrix}
			\chi' \e^{-i\theta} \\ c \\ \chi' c \e^{-i\theta} \\ 1
		\end{pmatrix}\allowdisplaybreaks[4] \\
		\psi_{b} &= b\e^{-ik'_{x}x}
		\begin{pmatrix}
			-\chi' \e^{i\theta} \\ c \\ -\chi' c \e^{i\theta} \\ 1
		\end{pmatrix}\allowdisplaybreaks[4] \\
		\psi_{t} &= t\e^{ik_{x}x}
		\begin{pmatrix}
			\chi \e^{-i\phi} \\ c \\ \chi c\e^{-i\phi} \\ 1
		\end{pmatrix}
	\end{align}
\end{subequations}
Here $r$, $a$, $b$ and $t$ are coefficients to be determined.
By matching the wavefunctions appropriately at $x=0$ and at $x=d$ (detailed in Appendix~\ref{app:finite_width_tunneling}), the tunneling probability is found to be:
\begin{widetext}
\begin{equation}
\label{eq:T_finite_trans}
T_{\mu} = \frac{\cos^{2}\phi\cos^{2}\theta}{\cos^{2}\phi\cos^{2}\theta\cos^{2}(k'_{x}d) + \sin^{2}(k'_{x}d)\left[1 - \mu\sin\phi\sin\theta\right]^{2}}
\end{equation}
\end{widetext}
where $\mu= \chi\chi' = +1$ denotes a same-chirality transmission and $\mu = -1$ denotes a chirality-flipping transmission.

\begin{figure}
	\includegraphics[width=6cm]{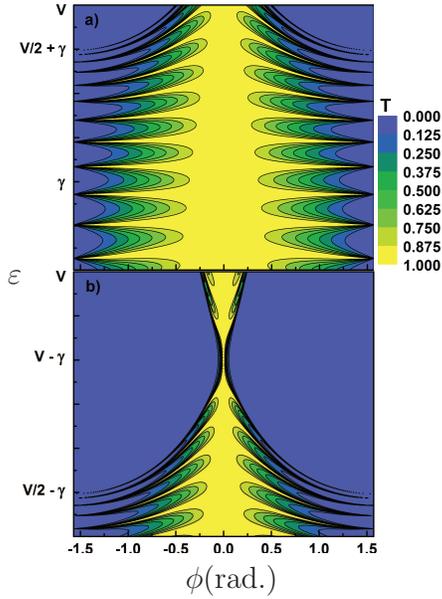}
	\caption{(Color online). Transmission probabilities for (a) upper and (b) lower cones across a $n$-$p$-$n$ junction.}
	\label{fig:npn_transmission}
\end{figure}

In the transmission probabilities shown in Fig. 5, Fabry-P\'{e}rot interferences are observed. Such interference patterns occur whenever a resonance condition is achieved within the barrier. Perfect transmission can occur for oblique incident angles whenever the condition $k'_{x}d = n\pi$ (where $n \in \mathbb{Z}$) is met. The incident energy for perfect transmission to occur at an oblique incident angle $\phi$ is given by
\begin{equation}
\label{eq:FP_full}
\veps_{n}(\phi) = c + \frac{\nu \pm\sqrt{\nu^{2}\sin^{2}\phi + \left(\frac{n\pi}{\beta}\right)^{2}\cos^{2}\phi}}{\cos^{2}\phi}.
\end{equation}
where $\beta = \frac{\gamma d}{\hbar\vf}$.

Although the transmission probability of the AA-stacked BLG has the same form as that of the SLG, there are several distinctions that can be made between the AA-stacked BLG and SLG. The most important distinction is the quasiparticle chirality. Since the band structure of AA-stacked BLG is made up of two shifted Dirac cones, the electrons and holes can both have $\chi=\pm1$ chiralities, whereas in SLG the chirality of an electron is always +1, and -1 for a hole. This difference manifests itself in the transmitted and critical angles which causes a deviance in the transmission probabilities (for both the $n$-$p$ and $n$-$p$-$n$ cases) from those of the single layer case. For the case of an $n$-$p$-$n$ junction in AA-stacked BLG, an incident $c=-1$ electron can be transmitted as either a $\chi = +1$ or a $\chi = -1$ intermediate hole state inside the barrier depending on the barrier height. For SLG with an $n$-$p$-$n$ junction, only the $\chi = -1$ state is available as an intermediate hole state \cite{klein_tunneling}. This chirality flipping behavior is only possible in AA-stacked BLG due to the up-down shifted Dirac cones.

\section{Junction Conductance}

\begin{figure}[h]
\includegraphics[width=6cm]{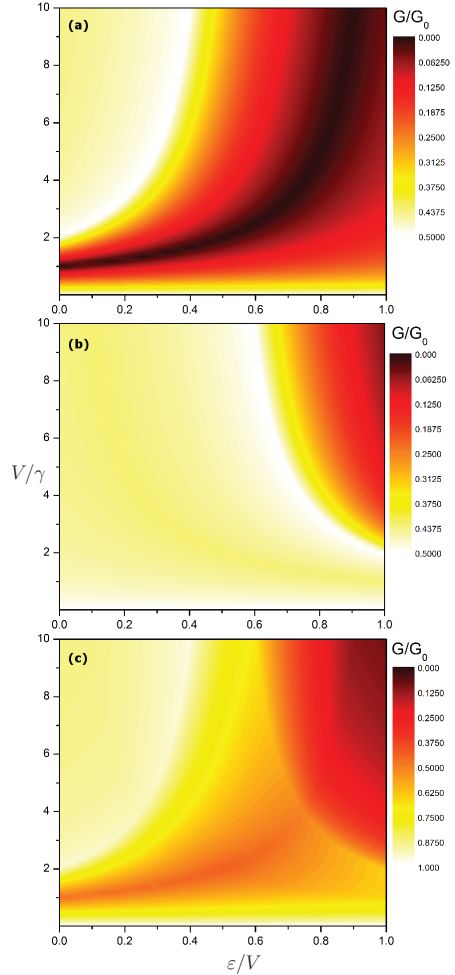}
\caption{(Color online). Conductance for (a) lower, (b) upper cones and (c) the conductance sum of upper and lower cones. The incident energy, $E$ is in units of $V$ and the potential height $V$ is in units of $\gamma$.}
\label{fig:conductances}
\end{figure}

The conductance across an $n$-$p$ junction of AA-stacked BLG is calculated in this section. We use the Landauer formula \cite{landauer_formula, angle_dependant_conductance} to obtain the conductance. The two terminal conductance $G(E,V)$ is written as:
\begin{align}
\label{eq:conductance}
&G(E,V) = 4\frac{e^{2}}{h} \sum_{ch}{T_{c}} \approx 2\frac{e^{2}W}{h\pi}\int^{k_{F}}_{-k_{F}}{T(k_{y})\dd k_{y}} \nonumber \\
&= \frac{8e^{2}Wk_{F}}{h\pi} \int^{\frac{\pi}{2}}_{0} {\frac{\cos^{2}\phi\sqrt{1-b^{2}_{c\mu}\sin^{2}\phi}}{1+\cos\phi\sqrt{1-b^{2}_{c\mu}\sin^{2}\phi} - b_{c\mu}\sin^{2}\phi}}dd\phi,
\end{align}
where `$c$' represents the two transmission channels through different cone, $W$ is the AA-stacked BLG sample width, and we define
\begin{equation}
b_{c\mu} = \frac{\gamma - cE}{\mu\left(V-E+c\gamma\right)}.
\end{equation}
Several special cases of Eq.~\eqref{eq:conductance} can be analytically determined:
\begin{subequations}
	\label{eq:isoconductances}
	\begin{align}
		&G_{c=\pm 1}\left(E =\frac{V}{2}\pm\gamma\right) = G_{0} \\
		&G_{c=+1}\left(E =\gamma\right) = (4-\pi)G_{0} \\
		&G_{c=-1}(E = V-\gamma) = 0 \\
		&G_{c=\pm 1}(V = 0) = G_{0}
	\end{align}
\end{subequations}
where $G_{0} = \frac{4e^{2}Wk_{F}}{h\pi}$ and $G_{c=\pm 1}$ is the conductance for the $c=\pm 1$ electrons. As the total conductance is defined as the sum of all transmission channels, the total conductance can be written as $G_{total} = (G_{c=+1} + G_{c=-1})/2$, meaning the conductances of each cone channel simply add to give the total conductance. A factor of $1/2$ is required since the total conductance is contributed by two cones. $G_{total}$ is integrated numerically and the results are shown in Fig.~\ref{fig:conductances}.

All the lines of equal conductance observed in the conductance plots for the upper and lower cones are expected by the analytically calculated values \eqref{eq:isoconductances}. These lines are curved, not straight as one would expect from the expressions due to the nature of the parameterization of the graphs.

We now discuss the differences in the conductance between AA-stacked BLG and SLG. 
For SLG, the incident energy can be scaled by the potential height, thus eliminating the potential height from the expression for conductance. For the AA-stacking however, the energy and potential are more naturally scaled by the inter-layer hopping energy, $\gamma$. This results in the conductance being a function of both incident energy and potential height, causing the conductance functions to be more complex in comparison to the single layer case. Furthermore, the total conductance for the AA-stacking is the (averaged) sum of the conductance contributed by two cones. The total conductance of AA-stacked BLG is therefore substantially different when compared to SLG.

The junction conductance for AB-stacked BLG is also different from AA-stacked BLG since the transmission probabilities of AB-stacked BLG \cite{chiral_tunneling, duppen} differ significantly from that of AA-stacked BLG. The tunneling conductance of AB-stacked BLG has been calculated by Duppen et. al. by using a four-band model \cite{duppen}. A major difference between the conductance of the AA-stacked and AB-stacked BLG is that the tunneling conductance in AB-stacked BLG is made up of four distinct components since transition between all four bands are possible. The onset of each transmission channel in AB-stacked BLG creates a discontinuity in the conductance plot. In contrast, the forbidden inter-cone transition allows only two transmission channels in AA-stacked BLG, resulting in a smoother conductance plot.

The total integrated conductance, $\mathcal{G}$, is determined by integrating Eq.~\eqref{eq:conductance} over the energy for each cone, i.e.
\begin{equation}
\mathcal{G}(V) = \sum_{cone}\int G(E,V)N(E)\dd E
\label{eq:total_conductance}
\end{equation}
where $N(E) = \frac{2\gamma}{\pi(\hbar \vf)^{2}}\left[\left|\frac{E}{\gamma} - 1\right| + \left|\frac{E}{\gamma} + 1\right|\right]$ is the density of states \cite{AA_conductivity} and the integral is taken over the energies of the incident electrons that contribute to the conductance.

\begin{figure}[h]
	\includegraphics[width=6cm]{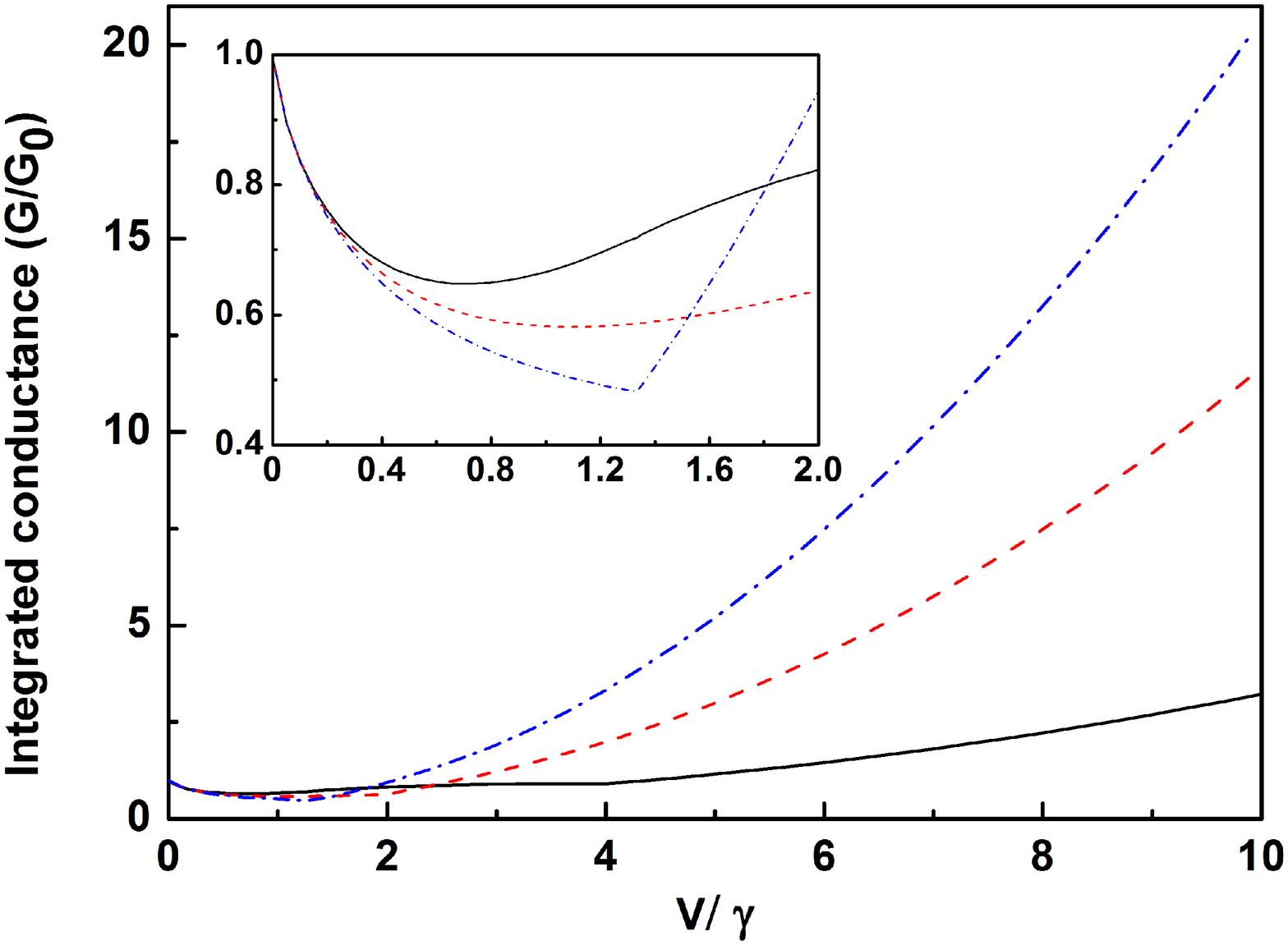}
	\caption{The total integrated conductance of an $n$-$p$ junction. The integrated conductances are plotted for three Fermi levels (in units of the potential height, $V$): $E_F/V=0.25$ (solid line), $E_F/V=0.50$ (dashed line) and $E_F/V=0.75$ (dash-dotted line).}
	\label{fig:total_conductance}
\end{figure}

The potential height dependence of the total integrated conductance is shown in Fig. 7 (Fermi energy $E_{F}$ is in the units of $V$). The limits for the integration varies between cones. Only the difference between the Fermi energy and $+(-)\gamma$ for the upper (lower) cone is required to be considered when integrating the conductance, i.e. only states lying above the Dirac point of each cone contributes to $\mathcal{G}$.
For example, if we consider $E_{F} > 0$, the limits of integration will be from $-\gamma$ to $E_{F}$ for the lower cone, and $E_{F}$ to $\gamma$ if $E_{F} < \gamma$, or $\gamma$ to $E_{F}$ if $E_{F} > \gamma$.
$\mathcal{G}$ exhibits a sharp turning point when the potential height (in units of $\gamma$) is equal to the inverse of the ratio of the Fermi energy to the potential. This occurs since the Fermi energy lines up exactly with the Dirac point with energy $\gamma$ and the transmission probability in this state is strictly zero. The upper and lower integration limits of Eq.~\eqref{eq:total_conductance} are both $\gamma$ in this case also. This results in the disappearance of the upper cone conductance, $G_{c=+1}$, and a sharp turning point occurs in the total conductance.

\section{Conclusion}

The cone conservation in electron transport reported here is only possible in the case of a gapless band dispersion. A finite electron-electron interaction can give rise to a gap opening at the Fermi level \cite{AA_stack_instability, prada}. In this case the cone conservation is not observed in the electron transport and both the reflected and transmitted states can be in any cone. This situation is very similar to single layer graphene where Klein tunneling is only possible when the band structure is gapless \cite{setare}. For AA-stacked graphene, it has also been shown that gap can close if the system is doped and the doping exceeds some threshold or the temperature is sufficiently high \cite{sboychakov1, sboychakov2}. In this case, the reported cone conservation is likely to be re-established as long as the electron Fermi level still lies in the linear dispersion regime.
The cone conservation is a direct consequence of Eq.~\eqref{eq:psi_ortho}. The total cancellation of the overlap of the wavefunctions when $c \neq c'$ occurs due to the simple form of the AA-stacked BLG eigenstates in which the components are only dependent on the angle of the incident wavevector. In the presence of a gap, such as the one introduced when an anti-ferromagnetic ordering is present\cite{sboychakov1}, the eigenstates become a far more complicated function of the wavevector, and the cancellation of the overlap between eigenstates of different cones is no-longer possible.

In conclusion, we have theoretically investigated the quantum tunneling of electrons in AA-stacked BLG. Since the band structure of AA-stacked BLG is made up of two shifted Dirac cones, Klein tunneling was found to occur in this system. The quasiparticles in AA-stacked BLG are not only chiral but also labeled by an additional cone index. Due to the orthogonality of the wavefunctions with different cone indices, the cone index is a conserved quantity. Together with the negative refraction of electrons, the protected cone transport across an AA-stacked BLG $n$-$p$ junction allows the transmitted electrons to be spatially separated according to their cone index. This suggests the potential of achieving `cone-tronic' devices in AA-stacked BLG. The junction conductances were calculated for an $n$-$p$ junction and a number of difference between the results for AA-stacked BLG and SLG were found. First, due to the shifting of Dirac cones in AA-stacked BLG, the conductance plots of each individual cone were shifted as compared to those of SLG. Second, since in AA-stacked BLG there are two transmission channels, the total conductance is the averaged sum of both channels in contrast to SLG where only a single cone is involved. In AB-stacked BLG, the total conductance is made up of four transmission channels whereas in AA-stacked graphene, only two intra-cone transitions are possible. Further, electrons are perfectly reflected at normal incidence in AB-stacked BLG and obey a considerably different tunneling probability to AA-stacked BLG. Finally, we purpose the theoretical results to be verified experimentally.

\begin{acknowledgments}
The work is supported in part by the Australian Research Council through a Discovery Project Grant (DP130102956).
\end{acknowledgments}

\appendix

\section{Derivation of transmission probability across finite width junction}\label{app:finite_width_tunneling}

By applying boundary conditions: $\psi_{i}(x=0) + \psi_{r}(x=0) = \psi_{a}(x=0) + \psi_{b}(x=0)$ and $\psi_{a}(x=d) + \psi_{b}(x=d) = \psi_{t}(x=d)$, a system of equations can be formed:
\begin{widetext}
\begin{equation}
\label{eq:general_system}
\begin{pmatrix}
\chi\e^{-i\phi} \\ c \\ c\chi\e^{-i\phi} \\ 1 \\ 0 \\ 0 \\ 0 \\ 0
\end{pmatrix}
=
\begin{pmatrix}
\chi\e^{i\phi} & \chi'\e^{-i\theta} & -\chi'\e^{i\theta} & 0 \\
-c & c& c & 0 \\
c\chi\e^{i\phi} & c\chi'\e^{-i\theta} & -c\chi'\e^{i\theta} & 0 \\
-1 & 1 & 1 & 0 \\
0 & \chi'\e^{ik'_{x}d}\e^{-i\theta} & -\chi'\e^{-ik'_{x}d}\e^{-i\theta} & -\chi\e^{ik_{x}d}\e^{-i\phi} \\
0 & c\e^{ik'_{x}d} & c\e^{-ik'_{x}d} & -c\e^{ik_{x}d} \\
0 & c\chi'\e^{ik'_{x}d}\e^{-i\theta} & -c\chi'\e^{-ik'_{x}d}\e^{i\theta} & -c\chi\e^{ik_{x}d}\e^{-i\phi} \\
0 & \e^{ik'_{x}d} & \e^{-ik'_{x}d} & -\e^{ik_{x}d}
\end{pmatrix}
\begin{pmatrix}
r \\ a \\ b \\ t
\end{pmatrix}
\end{equation}
\end{widetext}
Care needs to be taken with Eq.~\eqref{eq:general_system} as the transmitted angle, $\theta\;(\equiv\theta_{t})$, can have a phase of $\pi$ depending on the chirality index $\chi$ of the transmitted states (i.e. whether these states are electron-like or hole-like). Eq.~\eqref{eq:general_system} is in the form of $\tilde{y} = A\tilde{x}$ and hence we can find $\tilde{x} = A^{-1}\tilde{y}$ where $A^{-1}$ is the pseudo-inverse of $A$ given by $A^{-1} = \left(A^{\dagger}A\right)^{-1}A^{\dagger}$.
Only the reflection coefficient $r$ is of interest as the transmission probability can be calculated directly from $T = 1 - |r|^{2}$. It is found that the form of $r$ is dependent only on whether the electron transition changes the chirality. It is found that:
\begin{equation}
\label{eq:r_finite_trans}
r_{\mu} = \frac{-2\e^{-i\phi}\sin(k'_{x}d)\left[\sin\phi -\mu \sin\theta\right]}{\e^{ik'_{x}d}\cos(\phi-\theta) +\mu \e^{-ik'_{x}d}\cos(\phi+\theta) + 2i\mu\sin(k'_{x}d)},
\end{equation}
where $\mu=\chi\chi'=\pm 1$ indicates whether there is a sign change of the chirality index. The transmission probability $T_{\mu}$ is given as
\begin{equation}
\label{eq:app_T_finite_trans}
T_{\mu} = \frac{\cos^{2}\phi\cos^{2}\theta_{a}}{\cos^{2}\phi\cos^{2}\theta_{a}\cos^{2}(k'_{x}d) + \sin^{2}(k'_{x}d)\left[1 -\mu \sin\phi\sin\theta_{a}\right]^{2}}.
\end{equation}
\end{document}